# People-centric computing and communications in Smart Cities[1]

Franca Delmastro, Valerio Arnaboldi, Marco Conti
IIT Institute – National Research Council of Italy (CNR)
Via G. Moruzzi, 1
56124 Pisa – Italy
email: firstname.lastname@iit.cnr.it

**Abstract**

The extreme pervasive nature of mobile technologies, together with the users' need to continuously interact with her personal devices and to be always connected, strengthen the user-centric approach to design and develop new communication and computing solutions. Nowadays users not only represent the final utilizers of the technology, but they actively contribute to its evolution by assuming different roles: they act as *humans*, by sharing contents and experiences through social networks, and as *virtual sensors*, by moving freely in the environment with their sensing devices. Smart cities represent an important reference scenario for the active participation of users through mobile technologies. It involves multiple application domains and defines different levels of user engagement. *Participatory sensing, opportunistic sensing and mobile social networks* (MSN) currently represent some of the most promising people-centric paradigms. In addition, their integration can further improve the user involvement through new services and applications. In this paper we present SmartCitizen app, a MSN application designed in the framework of a smart city project to stimulate the active participation of citizens in generating and sharing useful contents related to the quality of life in their city. The app has been developed on top of a context- and social-aware middleware platform (CAMEO) able to integrate the main features of people-centric computing paradigms, lightening the app developer's effort. Existing middleware platforms generally focus on one single people-centric paradigm, exporting a limited set of features to mobile applications. CAMEO overcomes these limitations and, through SmartCitizen app, we highlight the advantages of implementing this type of mobile applications in a smart city scenario. Experimental results shown in this paper can also represent the technical guidelines for the development of heterogeneous people-centric mobile applications, embracing different application domains.

## 1  Introduction

The penetration of smartphones and wearable sensing devices in everyday life is driving toward the definition and the development of people-centric computing and communications paradigms [1] [2] . These are aimed at developing smart services able to increase citizens' participation and empowerment. Smart cities represent a relevant scenario for people-centric solutions in which, on the one hand, sensors and smart objects are used to monitor the city infrastructures (e.g., electric grid, lighting, transport, etc.) and, on the other hand, the users with their personal devices become *sensors* able to monitor the human behavior in the city and contribute to the definition of smart policies. Without a concrete user involvement, new policies for efficient energy consumption, smart mobility, traffic, environmental monitoring and others, are simply ineffective.

From a technological point of view, the user involvement can be achieved by leveraging three user-centric paradigms: (i) *participatory sensing [3]* in which the users exploit their sensing devices to collect information during their daily living activities and share them (with the other users) mainly through web servers; (ii) *opportunistic sensing [5]*, in which mobile applications opportunistically exploit all the sensing technologies available in the environment (not requiring a direct user interaction), and (iii) *opportunistic Mobile Social Networks (MSN)* [4], in which the users directly generate and share heterogeneous types of contents with nearby users in real-time, by exploiting the physical interactions of their personal mobile devices (i.e., opportunistic communications). The emergence of these three paradigms witnesses a substantial change in the management of legacy ad hoc and sensor networks, creating new research challenges related to sensing devices, wireless communications, data management and dissemination.

By integrating the features of these user-centric paradigms in a single mobile system, we can design efficient and personalized mobile applications in several domains (e.g., health and well-being, social inclusion, smart mobility, environmental and urban monitoring), and integrate them towards the real development of a smart city.

To this aim, we designed and developed CAMEO [6], a context- and social-aware middleware platform for mobile devices and MSN applications. It provides app developers with the functionalities to automatically discover users and mobile devices nearby, and to identify users' common interests, habits, available services, and resources. All this data is collected and exchanged through opportunistic communications and it represents the *context* used by CAMEO to implement optimized networking protocols, resource management mechanisms, and to export context-aware features to MSN applications. CAMEO is able to support concurrent MSN apps on the same device through dedicated inter-process communications, and it is deployed as an Android application running in background.

As a real example of MSN app running on top of CAMEO, we developed SmartCitizen app. It has been designed in the framework of the research project SmartHealthyENV [2] to involve citizens in collecting and sharing personal experiences about the quality of life in the city of Pisa (Italy). The main topic of discussion in SmartCitizen is related to the city environmental conditions (e.g., air quality, weather, pollution) since one of the project's targets is the development and deployment of a low-cost sensing infrastructure for environmental monitoring. The idea is to demonstrate that a medium-scale city can exploit a low-cost sensing infrastructure to improve the current environmental monitoring coverage enriched with user-generated contents, both objective (e.g., picture or videos of specific locations), and subjective, (like comments, suggestions, remarks). In this architecture, environmental sensor data are continuously collected and periodically transmitted to a centralized server implementing Sensor Web Enablement (SWE) standard[3], the reference standard for IoT and Web of Things scenarios. Generally, mobile devices are not able to process SWE data due to the complexity of XML schemas used in the standard, and to the lack of efficient software tools for processing them [7] CAMEO overcomes this limitation by integrating Sensor Mobile Enablement (SME) [8] , a software library implementing a set of light-weight standard models compliant with SWE (including also general schemas for phone-embedded sensors) and efficiently manageable on mobile devices. SmartCitizen, therefore, can interact with external SWE services, generally dedicated to environmental monitoring, and then exploit opportunistic communications to implement opportunistic sensing features, like forwarding environmental data to users and devices without Internet connectivity, or share its own sensing resources with the others on demand. Finally, SmartCitizen implements the participatory sensing features through the generation of user contents that are subsequently disseminated by CAMEO based on common users' interests, habits or similar profiles.

---

[2] Smart HealthyENV -Smart Monitoring Integrated System for a Healthy Urban ENVironment in Smart Cities, project funded by Tuscany region (Italy) under the program POR-CREO FESR 2007-2013, http://www.progettoshe.it

Through SmartCitizen, we have been able to evaluate pros and cons of MSN applications, both from the users' perspective, through experimental evaluation, and from the technical point of view, in terms of development of top of CAMEO, the efficient management of data derived from heterogeneous sources and services and the use of the novel WiFi Direct standard to deploy opportunistic communications and content dissemination protocols. This represents an important feedback for mobile app developers willing to integrate heterogeneous people-centric paradigms in a single mobile app.

The paper is organized as follows. In Section 2 we present in detail SmartCitizen features. Section 3 shows the experimental evaluation divided into three different subsections related to heterogeneous context management, user experience and opportunistic communication issues, respectively. Finally, Section 4 presents the advantages of the proposed solution and of the entire framework with respect to existing work and future research directions.

## 2 SmartCitizen App

SmartCitizen is aimed at stimulating the active participation of citizens in collecting and sharing data related to the environmental conditions of the city of Pisa. The first approach to the app is represented by an intuitive visualization of the current environmental conditions of some city areas, used as testbed for the monitoring infrastructure developed inside SmartHealthyENV project. Then, the app stimulates the creation of discussions between citizens on topics related to air quality by borrowing functions from social networking applications, such as posting, commenting and chatting. Of course, discussions are not limited to environment-related topics, and citizens can use the app as a traditional mobile social network. In Figure 1 we present some screenshots of the graphical user interface of the app showing a real use case scenario. In the first screenshot, the app presents on a map the air quality index measured on each sensing station deployed in the city and presented as a circle of a range of colors from red (polluted area) to green (healthy area) for an intuitive comprehension by common citizens. However, the app is designed also for expert users, such as employees of the local municipalities, able to interpret the detailed information about measured chemical parameters. In this case, the authorized user, clicking on the circle, will visualize the detailed data of the related station. All the sensing data is downloaded from the SmartHealthyENV server if the user device has Internet connectivity, or it can be downloaded through device-to-device communication if available on other user devices in proximity. As far as the generation of user contents, Figure 1 shows the different views provided to the user: the form for the creation of a new post, the list of active discussions on different topics, and the list of generated content for each discussion with appropriate notification of new available contents. In the use case shown in Figure 1, users commented about the weather in a specific area generally used as pedestrian path, supported by the green air quality index in the same area and by some multimedia content. To further customize SmartCitizen for specific user categories, we integrated some features for sports users (Figure 2), who are generally interested in environmental conditions for their outdoor activities, they prefer making activities in healthy areas, and most of them use to compare their experiences and performances through dedicated or online social networks. In this case, SmartCitizen allows users to directly record activity paths in the city, it associates the air quality information derived from the monitoring network with each path and the user can further customize it with her own information, comments and suggestions. These features allow users to share immediate information about the specific paths and areas in the city.

---

[3]OGC Sensor Web Enablement DWG (SWE), http://www.opengeospatial.org/projects/groups/sensorwebdwg

All the contents generated though the app are disseminated on the network through the context-aware content dissemination protocol provided by CAMEO. To this aim, the app developer must only define the application user profile: a set of information characterizing the user in the framework of the current application (e.g., user category, interests related to generated contents and/or specific application features). This profile is automatically shared among mobile devices in proximity running CAMEO, and it is used to evaluate the potential utility of the available contents provided by other users in the network for the local user. To this aim, each user should characterize the generated content (generally multimedia contents like audio/video messages, often associated with text posts) by selecting or generating one or more tags, which are used as input for the matching "user-content" algorithm. Since users' interests can change over time, CAMEO automatically adapts the dissemination protocol to those changes, maintaining all the procedure transparent to the app developer.
Users' profiles and contents are distributed on the network through peer-to-peer communications. Nodes communications happen through WLAN access or through a WiFi Direct self-organizing network, based on the current availability. Also in this case, the app is completely transparent to the type of communication protocol used, which is automatically selected by CAMEO.

In order to enrich the centralized storage of environmental data with user–generated content, SmartCitizen allows also users to upload their contents to SmartHealthyENV server, maintaining the compliance with SWE data format. This is not automated to limit the connections towards the server and to push the user to upload only significant contents.

In terms of privacy, users can decide what type of information to share in proximity and what can be stored remotely. They must explicitly agree to share them before starting the app through a specific privacy statement. However, no personal and sensible data are exchanged with the others.

In the next Section we present the experimental evaluation of SmartCitizen in terms of: *(i)* efficient management of heterogeneous sensing data based on SWE standard; *(ii)* user experience in a real setting, and *(iii)* open issues related to opportunistic communications based on WiFi Direct.

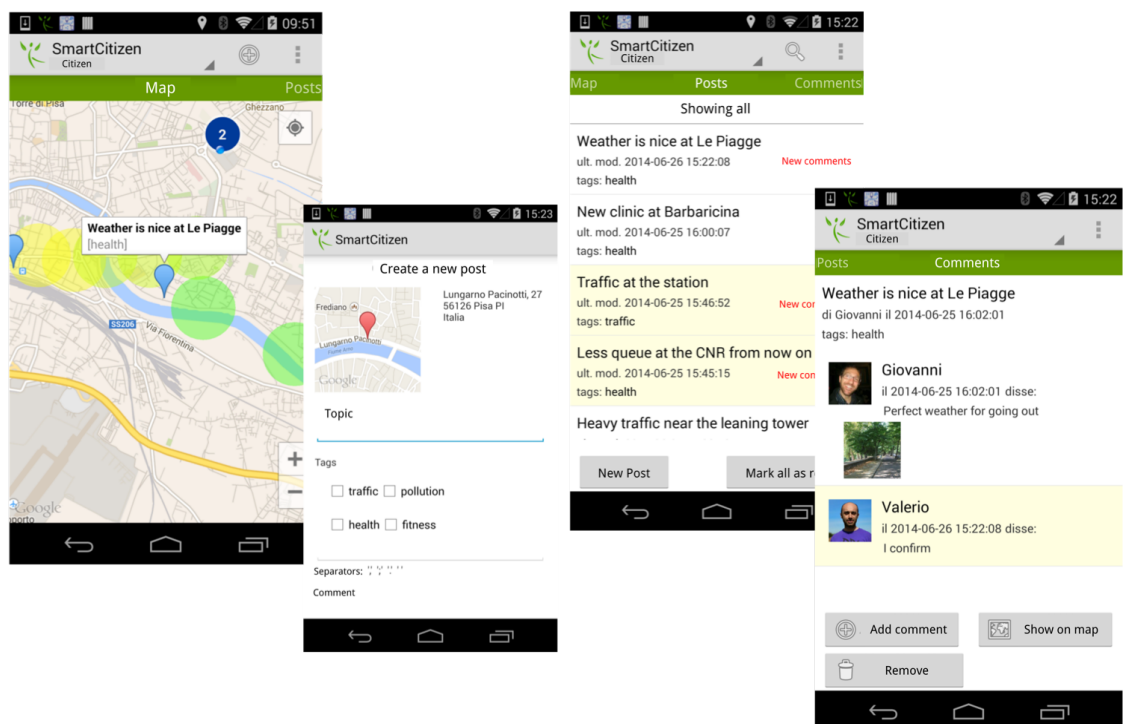

**Figure 1** SmartCitizen use case scenario.

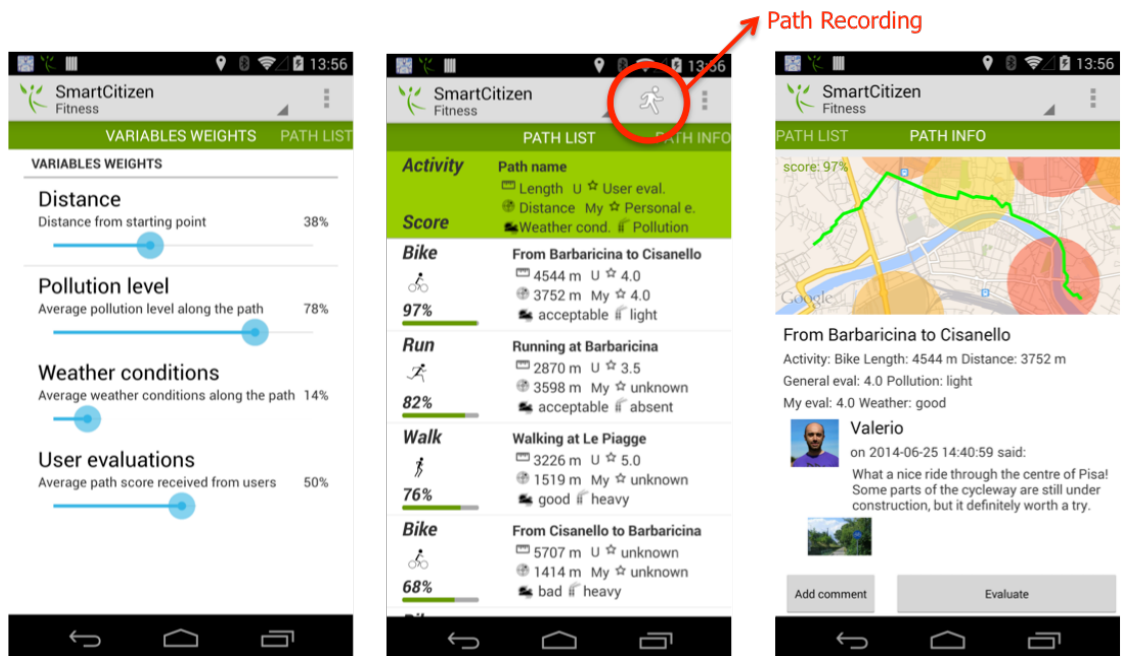

Figure 2: SmartCitizen for Fitness

## 3 Experimental Evaluation

### 3.1. Heterogeneous sensor data management

To perform efficient data exchange with SmartHealthyENV server, the app uses SME encoding and decoding procedures integrated in CAMEO. SME has two main targets:
- (i) to define the descriptions of phone-embedded sensors and related measurements compliant with SWE standard;
- (ii) to efficiently process SWE sensor data on mobile devices.

The interaction with a SWE server requires a series of steps, as defined by the standard. The application must first query the server to obtain a description of the monitoring service in term of available sensors and their properties (e.g., vendor name, sampling frequency, range of valid measurements). Then, it can ask for the measurements of single or multiple sensors (called Sensor Observations). In case of citizen users, the app asks only for the indexes as the elaboration of multiple raw parameters on the same sensing station. Instead, for an expert user, it can request for the detailed data on a variable time window, in order also to visualize temporal variations of the single parameters. Therefore, the size of the SWE files downloaded from the server can vary depending on the application request, and it can impact on the mobile device processing time for data serialization and deserialization procedures.

For this reason, to evaluate the efficacy of SmartCitizen in managing heterogeneous sensor data through SME, we performed several experiments in a real environment. Specifically, we measured serialization and deserialization times of a SWE file containing a variable number of raw sensor measurements.

The file contains an increasing number of values related to independent measurements (from 0 up to 50,000 with steps of 10,000). This represents a complex and large SWE file that can be received by a mobile device (for which SWE deserialization is required) and subsequently forwarded from a mobile device to another (for which SME serialization is required before transmission).

The XML file size of this observation ranges between 1.9KB (i.e, containing only SWE XML header and 0 values) and 2.3MB (i.e., containing 50,000 values). Results shown in Table 1 highlight the efficiency of SME library to manage this operation in few seconds [4]. Specifically, deserialization and serialization times remain reasonably small even in case of 50,000 sensor observations, which represent a high number of observations for a limited area, such as the neighborhood of a single user in an urban scenario.

| # values in a single observation | SWE XML File size | Avg Deserialization Time (sec) | 95% confidence interval | Avg Serialization Time (sec) | 95% confidence interval |
|---|---|---|---|---|---|
| 0 | 1.9KB | 0.019 | 0.005 | 0.101 | 0.049 |
| 10000 | 460KB | 0.314 | 0.017 | 0.891 | 0.064 |
| 20000 | 917KB | 0.644 | 0.015 | 1.534 | 0.125 |
| 30000 | 1.4MB | 1.181 | 0.058 | 2.318 | 0.258 |
| 40000 | 1.8MB | 1.290 | 0.052 | 2.961 | 0.367 |
| 50000 | 2.3MB | 1.513 | 0.071 | 3.864 | 0.396 |

**Table 1** Serialization and deserialization times of a single SWE observation with an increasing number of values.

### 3.2. User experience

Before releasing the app on the store and open the test to citizens, we performed an experiment to assess the user experience in real settings. We involved 19 participants in a one-day session for testing the app, analyse reactions and collect suggestions. The participants were 17 students and 2 professors of a high school class who planned a visit to our campus to learn about new technological solutions for smart cities. We briefly presented the research project and the features of SmartCitizen, and we asked them to install CAMEO and the app on their Android smartphones. Then, they could use the app while moving around the campus in order to generate contents and communicate with each other. The experiment lasted three hours (excluding the initial training session). At the end of the visit, we asked the users to fill in a questionnaire designed to identify their technological skills and to evaluate their experience while using the app.

The majority of the users declared medium-high skills in the use of new technologies (i.e., all of them spend at least one hour per day on Internet and they all use OSN applications like Facebook or Twitter). They also understood very quickly how to use the paradigm of MSN inside the app, as demonstrated by the amount of generated and shared content, as detailed in Table 2.

---

[4] In this case the difference between serialization and deserialization times is mainly due to the different time cost of read and write operations on the device mass memory.

During the experiment, we personally contributed generating a couple of posts just to show to the participants how to use the app. Then, they actively involved us as additional users in the experiment, adding comments to our posts. Therefore, the effective number of people involved in the experiment was 22, creating, in total, 39 posts, 151 comments and 22 photos attached to posts or comments. These contained also 22 additional tags (i.e., other than those already pre-defined by the application), customizing then the content dissemination protocol on new user-defined interests. Out of the 22 participants, 13 created new posts during the experiment, 20 generated at least one comment to an existing post, 13 generated both posts and comments, and no one generated only comments. Moreover, 8 users created multimedia content (photos or videos) in their posts or comments. The statistics of the generated contents in terms of posts, tags, and photos are reported in Table 2.

| Measure | Average | SD | Min | Max |
|---|---|---|---|---|
| posts per user | 1.77 | 2.18 | 0 | 8 |
| tags per post | 1.72 | 1.19 | 1 | 5 |
| comments per user | 6.86 | 5.58 | 0 | 21 |
| comments received per post | 4.21 | 4.33 | 1 | 22 |
| tags created per user | 2.32 | 3.01 | 0 | 12 |
| posts per tag | 2.9 | 3.84 | 1 | 15 |
| comments per tag | 11.52 | 17.24 | 1 | 57 |
| photos per post | 1.22 | 0.73 | 1 | 4 |
| photos per tag | 1.78 | 1.11 | 1 | 5 |

**Table 2: Summary of data generated during the experiment.**

Considering the limited time of the experiment and the concurrent visit to our campus, experimental results indicate that the participants have been active in the generation of content, and they immediately felt comfortable with the app usage. Even though the number of posts per user may seem rather small, it is worth noting that the users created a large number of comments, as each post generated threads of communication among the participants.

Through the subsequent questionnaire, to evaluate the user experience while using the app, we asked them to assess the quality of the app by assigning a score to the following fields: ease of use, usefulness, ease of content creation, and graphical quality. The results, depicted in Figure 3, highlight that the user experience was, by and large, positive ("medium-high" quality). For the ease of use and ease of content creation, some of the "mid-low" evaluations (2 out of 5) came from participants who self reported low skills in using new technologies. Instead, the mid-low evaluations about the usefulness of the app might be negatively affected by the execution of the experiment in a controlled environment and for a limited amount of time. However, the users declared their interests in downloading and using the app also outside the experiment once the app will be released on Google Play store. In fact, it is worth noting that currently the app is still in a prototype version, and some of the negative evaluations might be due to minor glitches that the participants found during the experiment.

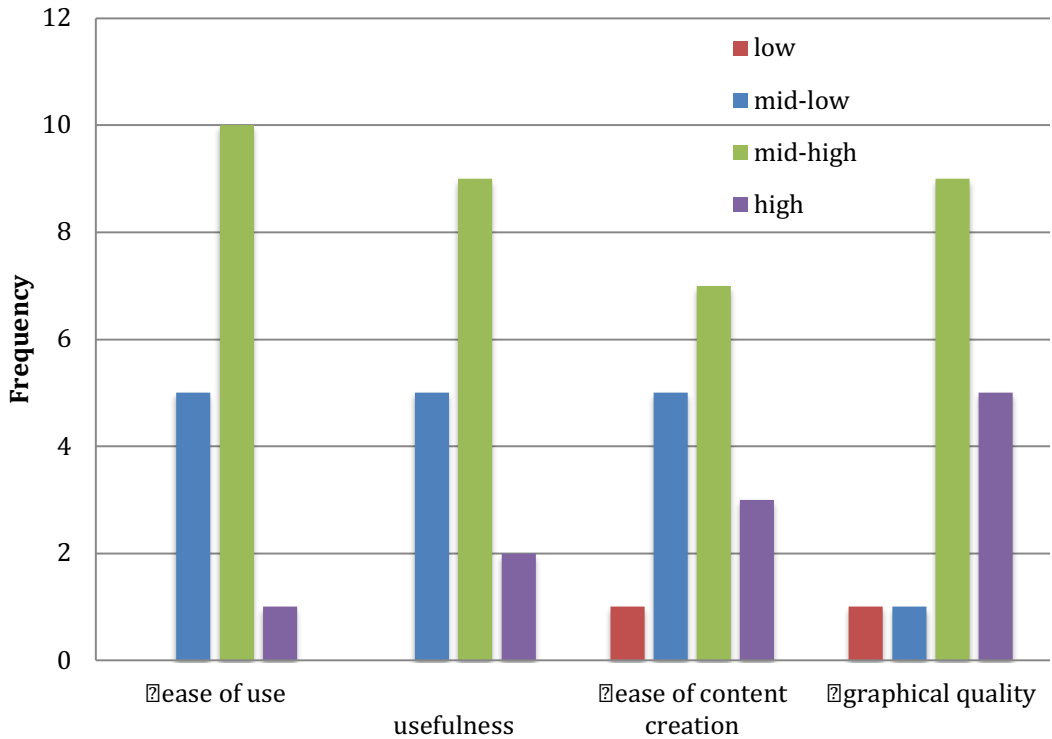

**Figure 3**: User experience evaluation based on questionnaires.

### 3.3. Open issues in WiFi Direct Opportunistic communication in real environments

As previously described, SmartCitizen has no control over the type of communication protocol used to disseminate data over the network, depending on the availability of a WLAN access. Actually, the deployment of real opportunistic networks based exclusively on device-to-device (D2D) communications is still an open research challenge, especially trying to use new emerging standards that should effectively allow the set up of large-scale testbeds by directly using heterogeneous user devices.

WiFi Direct currently represents the most promising technology to support D2D in opportunistic networks and it is supported by CAMEO. The main feature of WiFi Direct is the creation of *groups* of nodes (peers) within which each node can communicate with the others. A single node takes the role of *Group Owner* (GO), acting as an infrastructure WiFi BSS, and lets other nodes in proximity, called *clients*, to establish a connection with it. Communications are automatically cyphered at the network level within WiFi Direct groups, and a password is automatically exchanged between the peers of a group through an initial handshake phase. Users must only accept the wireless connection with the other device at the first meeting, like in Bluetooth pairing procedure.

Considering the amount of data exchanged in SmartCitizen through WiFi Direct, we noticed that we have limited traffic load but there can be issues related to the group management and to the self-organization of the network in case of nodes' mobility and intermittent connectivity conditions, which also represent the main characteristics of an opportunistic network. Here, we present the results of a series of experiments that we conducted to evaluate WiFi Direct performances in terms of throughput and content dissemination, considering those as essential features for a large-scale deployment.

| Group size | GO↔client | Client↔client | 1) GO↔client<br>2) client↔client | 1) Client(B)↔client(A)<br>2) Client(C)↔client(A) | 1) GO↔client(A)<br>2) Client(B)↔client (A)<br>3) Client(C)↔client(A) |
|---|---|---|---|---|---|
| 2 | 54.4Mbit/s | - | - | | - |
| 3 | 52.6Mbit/s | 22.3Mbit/s | 1) 44.3Mbit/s<br>2) 4.24Mbit/s | | - |
| 4 | 52.75Mbit/s | 17Mbit/s | 1) 40Mbit/s<br>2) 5.41Mbit/s | 1) 12.7Mbps<br>2) 9.07Mbps | 1) 37.4Mbit/s<br>2) 2.9Mbit/s<br>3) 3.22Mbit/s |

**Table 3 Throughput in WiFi Direct groups with 2, 3, and 4 peers. A↔B indicates that the throughput is the average for unidirectional communication from A to B and from B to A.**

WiFi Direct standard does not define a limit to the number of clients of a group even though the star configuration can limit the network performances with a high number of nodes. In fact, even with few nodes, we experimentally evaluated that the throughput drastically decreases when the number of client-to-client communications within a group increases. Specifically, we measured the throughput in a WiFi Direct group with a variable number of peers that communicate with each other for 180 minutes. In Table 3, we present different traffic configurations (e.g., GO-client, client-client, etc.) for different group sizes (2, 3 and 4 nodes). In the first configuration we created a group of two peers, a GO and a client, and we generated TCP traffic from the former to the latter (and vice versa). The average throughput after an experiment of 180 minutes of continuous traffic was 54.4Mbps. This represents a rather high throughput, supporting also the exchange of large files in short time periods. This makes the technology suitable for opportunistic scenarios in which the high mobility of nodes limits the duration of physical encounters and, consequently, the time opportunity to exchange data, like the urban scenario of SmartCitizen.

In a scenario with three peers per group, considering the communication between the GO and one client, and vice versa, we obtained broadly the same results of the previous case in terms of throughput. Despite this, communications from client to client lead to a throughput value that is nearly half the previous case. This is due to the fact that client-to-client communications require two steps before reaching the destination, forcing the communication to pass through the GO. To better investigate the limitation of the star topology in an opportunistic network, we set up an experiment with multiple client-to-client communications within groups of four peers. In this case, considering the worst-case results (i.e., the GO and two clients simultaneously send data to a single client), the throughput related to the GO-client link is around 37Mbps, while that obtained for the client-to-client links is around 3 Mbps. By considering these results, it is clear that client-to-client communications are penalized by the group size and this aspect has to be taken into account while designing forwarding and content dissemination protocols in opportunistic networks. Specifically, the framework could favour small groups with high reconfigurability based on users' mobility.

In addition, WiFi Direct standard works on a dedicated WiFi interface and this allows mobile devices to use both WiFi standard and WiFi Direct interfaces in specific configurations. For example, two devices belonging to two different WiFi

Direct groups can set up an inter-group communication, allowing thus the creation of a *multi-group*. In this scenario, one of the clients of a group is connected to the GO of a second group through its standard WiFi interface. This peer acts as a bridge that connects the two groups and enables the exchange of data between them. However, only the bridge is able to receive and send data from/to the other group. In fact, in order to forward contents from one node to all the others belonging to the two groups, we should implement a customized routing protocol (in charge of maintaining also information about the physical configuration of each group) as described in [9] and [10] .

Despite these limitations, we decided to investigate the feasibility and advantages of using multi-groups in our scenario, by conducting some experiments with particular attention to throughput analysis. Specifically, we measured the average throughput considering two concurrent communications from two clients of the first group to the client of the second group (connected through a dedicated interface to the GO of the first group. In this scenario the average throughput is 6.8Mbps, comparable with the results found in simple groups with the same cardinality. This means that the use of an additional interface and a separate channel between groups does not impact on the data traffic. However, Android OS does not support the automatic creation of multi-groups, since each group has a private key used to encrypt the communication, and there is no way to exchange such a key between members of different groups (neither between the GOs). Therefore, only manual configuration is possible. This makes multi-groups suitable only for customized experiments, but clearly not usable in real use case scenarios. For this reason, and for the purpose of our work, we decided to use only independent WiFi Direct groups, and to avoid the creation of multi-groups.

## 4    Related and Future Work

In the last few years, different middleware solutions have been proposed in the literature to support participatory sensing, opportunistic sensing, and mobile social networks. Despite this, to the best of our knowledge, all these solutions operate independently, generating thus several independent mobile applications. The integration of these paradigms in a single mobile platform is still to be achieved and, consequently, the definition of a single mobile application model that exploits all these features. Through CAMEO we intend to fill this gap in order to support a much larger range of people-centric services than existing middleware solutions, and SmartCitizen is an example of how to make it real.

We can list several works on the three single paradigms cited above, each of them focused on the improvement of a specific feature. As far as participatory sensing is concerned, Phokas et al. [11] have recently proposed a middleware aimed at simplifying applications' access to devices' sensors. MobIoT [13] is designed to monitor the sensing capabilities of mobile devices over time, trying to limit the amount of redundant data, temporarily deactivating redundant sensors and saving resources. Usense exploits sensors on mobile devices distributed in the environment to perform community-driven sensing tasks.

In the literature there are also middleware implementing device-to-device communications to support opportunistic data exchange between mobile devices, some of them based on the definition of context information to improve data dissemination, as demonstrated in Haggle project [15] . Finally, moving towards the application layers and especially referring to their social aspects, Yarta [14] is one of the few middleware platforms that supports the development of social-aware applications introducing the concept of Mobile Social Ecosystems (MSE), the set of interactions occurring among users and devices in a specific physical location. However, Yarta considers each MSE as an independent entity, not analyzing possible correspondences with users belonging to different ecosystems, and without any historical management of the context information, which can further profile the user.

As presented in [6] , CAMEO overcomes both Haggle and Yarta in the management of context, including a much more extensive context definition. In addition, it represents a step ahead with respect to previous solutions by integrating participatory and opportunistic sensing paradigms, and providing app developers with a variety of context- and social-aware features. In this paper we evaluated the effective advantages of using CAMEO as single middleware through the development of SmartCitizen. We analyzed both the technical performances of the app and the user experience, presenting positive feedback. In order to extensively evaluate the entire framework in a real environment, we are planning in the next future to introduce additional features able to support the deployment of large-scale experiments. In this way, we will be able to directly involve citizens in a real smart city scenario.

## Authors' short bio

Franca Delmastro is a Researcher at the IIT Institute of the Italian National Research Council. Her main research interests are in the design and development of innovative ICT solutions for smart cities, with particular attention to the health and well-being domain. Those solutions space from networking protocols for wireless and body area networks up to context- and social-aware middleware and mobile applications. She published several papers in international journal and conferences. She is an Editorial board member of International Journal of Communications, Network and System Sciences (IJCNS), and she participated in the organization of several international conferences.

Valerio Arnaboldi is a researcher at the IIT Institute of the Italian National Research Council. He works in the area of the analysis and characterization of social structures and information dissemination in Online Social Networks, and on context- and social-aware middleware solutions for Mobile Social Networking applications. He received a Ph.D in Computer Engineering from the University of Pisa in 2014. He co-authored the book "Online Social Networks: Human Cognitive Constraints in Facebook and Twitter Personal Graphs".

Marco Conti is a Research Director of the Italian National Research Council. He has published 300+ research papers, and five books, related to computer networks, pervasive systems and social networks. He received the Best Paper Award at IFIP TC6 Networking 2011, IEEE ISCC 2012 and IEEE WoWMoM 2013. He is Editor-in-Chief of Computer Communications and Associate Editor-in-Chief of Pervasive and Mobile Computing. He served as TPC/general chair for several major conferences, including: Networking 2002, IEEE WoWMoM 2005 and 2006, IEEE PerCom 2006 and 2010, ACM MobiHoc 2006 and IEEE MASS 2007.